\documentstyle[12pt]{article}
\begin{document}

\begin{center}
{\large BATALIN-TYUTIN QUANTISATION OF THE SPINNING PARTICLE MODEL }\\
\vskip 2cm
Subir Ghosh\\
\vskip 1cm
Physics and Applied Mathematics Unit,\\
Indian Statistical Institute,\\
203 B. T. Road, Calcutta 700035, \\
India.
\end{center}
\vskip 3cm
{\bf Abstract:}\\
The Spinning Particle Model for anyon is analysed in the Batalin-Tyutin
scheme of quantisation in extended phase space. Here additional degrees
 of freedom are introduced in the phase space such that all the constraints
in the theory are rendered First Class
that is commuting in the sense of Poisson Brackets. Thus the
theory can be studied
without introducing the Dirac Brackets which
 appear in the presence of non-commuting or Second Class constraints.
In the present case the Dirac Brackets make
the configuration space of the anyon non-canonical
and also being dynamical variable dependent, poses problems for the
 quantisation programme.
 We show that previously
obtained results, (e.g. gyromagnetic ratio of anyon being 2), are recovered
in the Batalin-Tyutin variable independent sector
in the extended space. The Batalin-Tyutin
variable contributions are significant
and are computable in a straightforward manner. The latter can be
understood as manifestations of the non-commutative space-time
in the enlarged phase space.

\newpage

\begin{center}
{\bf I: Introduction}
\end{center}
\vskip .2cm

Theoretical models for anyons \cite{1}, excitations  of arbitrary spin 
and statistics 
in 2+1-dimensions, have been proposed from different 
perspectives, the most popular ones being the point charge-
Chern-Simons gauge field interacting theory \cite{2} and the later 
Symplectic Models (SM) \cite{3,4} and the Spinning Particle Models 
(SPM) \cite{5,6}. The SM
and SPM can be related to the anyon field equation proposed in \cite{7}.
 It has been questioned \cite{8} whether the former interacting theory is a 
minimal description of the anyon. The latter models are free from 
complications of this type and both of them agree regarding results,
 (such as rigidity of the angular momentum and the gyromagnetic ratio
 $(g)$ being equal to $2$),
for free anyons as well as anyons interacting with abelian external
 gauge field, respectively.

However, the SM or SPM's are also interesting from another point of 
view. They have a close connection with the non-commutative space-time
theories, which have created  a lot of interest lately \cite{9}.

In SM \cite{4}, the symplectic structure has been posited, which gives 
rise to
the non- commutativity in the particle configuration space. On the other
hand, in SPM \cite{5,6}, the above mentioned feature is a result of the 
constraint
structure of the model. The set of Second Class Constraints (SCC),
 with non-commuting Poisson Brackets (PB), (in 
the sense of Dirac classification scheme \cite{10}), induces non-trivial
Dirac Brackets (DB) among the coordinate variables. Since we will 
concentrate on SPM \cite{6} in the present work, these comments will be 
elaborated.

Let us now impose a bit of caution on 
the methodology of evaluation of the $g$ value
of anyon in SM and SPM. One compares the Hamiltonian 
(in the former \cite{4}) and
equations of motion (in the latter \cite{6}) with the corresponding 
expressions
obtained for a charged particle with spin in backgroung electromagnetic
field {\it in conventional (commuting) configuration space}. The validity
of this matching procedure can be questioned on the grounds that the 
nature of the configuration spaces in the 
two systems that are being compared are
qualitatively different. But more
importantly, the quantisation procedure runs into trouble due to
operator ordering ambiguities in the DB's in SP and SPM. This complication
 arises since in theories with non-linear constraints, the DBs can involve
 dynamical variables
  which is true for the present case. When these DBs are
   elevated to commutation relations
 in the process of quantisation, the above mentioned feature creates problems.
 
 Finally, 
(as has been pointed out before \cite{5,6}), the induced non-vanishing 
 DB 
algebra $[A_a({\bf x},t),~ A_b({\bf y},t)]$  between supposedly "external"
 electromagnetic field variables makes us wonder as to how
far the treatment of the gauge fields being non-dynamical is justified.
 The problems related to the non-canonical nature of the coordinate 
 system have also been discussed in \cite{11}.

Now we come to the motivation of the present work. All of the above
problems can be naturally addressed in the Batalin-Tyutin (BT) quantisation
scheme \cite{12}, (which is a particular type of the more general 
quantisation
method \cite{13}), where additional phase space degrees of freedom, 
 (BT variables), are introduced in such a way that existing SCC's are
converted into  (abelian) First Class Constraints (FCC, see \cite{10}), thereby
enhancing the gauge symmetry of the extended system. Thus the problematic
DB's can be avoided. One can resort to Dirac quantisation of FCC system
by requiring that the physical states obey the FCCs, (i.e. $FCC\mid 
physical ~state\rangle =0$). Or else one can work in a convenient gauge where
the problematic DBs do not appear. Later on we will use the first
alternative.

Apart from the above reasons, we will find that by its own right,
 the problem of BT extension of SPM has several interesting features, 
some of them not addressed in the literature. Firstly, the  SCC system
in SPM can be expressed \cite{5,6} in a  covariant (and neat) form 
which however
is reducible that is the SCCs are not independent. The BT formulation
has been developed for irreducible or independent set of SCCs only. This forces
us to choose an independent set of SCC thereby losing manifest
covariance. As un unsolved problem, it will be interesting to study the BT
extension of a generic reducible SCC system.

Let us put the present work in its proper perspective. We have 
developed a framework for treating the Spinning Particle Model as
an FCC system in a BT-extended phase space where the background
space-time is normal, (that is commuting), so that comparison
between results obtained for anyons and analogous 
 models in conventional space-time 
do not pose any ambiguity. (In particular, we will recover the result
that for anyons $g=2$ but will indicate the presence of correction
 terms as well.)
Also the canonical quantisation programme 
is rigorous since DBs do not appear and so operator ordering
problems are absent in the canonical commutation relations.

Lastly we note that the present BT extension is classical in the sense that
operator ordering problems in the extended system have not been 
addressed \cite{14}.

The paper is organised as follows: in section II we briefly reproduce
the skeleton of BT formalism. This will also help us to fix the notations.
We quickly introduce the SPM in section III for completeness. Section IV
is devoted to developing the BT extension of SPM. This constitutes the 
main body of our work. Section V deals with the quantisation procedure 
and recovering of previous results. In section VI we conclude with a
discussion and some future lines of work.
\vskip .5cm
\begin{center}
{\bf II: Batalin-Tyutin Formulation}
\end{center}
\vskip .2cm
In this section we state the main results of BT 
prescription \cite{12} to be used. The basic idea behind the scheme is to 
 introduce additional phase space variables (BT variables) $\phi ^\alpha _a $,
 besides the existing physical degrees of freedom $(q,p)$, such that {\it all}
 the constraints in the extended system are reduced to FCCs. This means 
that one has to modify the original constraints and Hamiltonian 
accordingly by putting BT-extension terms in them. The way to achieve this 
at the classical level has been provied in \cite{12}.
Let us consider a set of constraints $(\Theta _\alpha ^a,\Psi _\l )$ and
a Hamiltonian operator $H$ with the following Poisson Bracket (PB)
relations,
$$
\{\Theta ^a_\alpha (q,p) ,\Theta ^b_\beta (q,p)\}\approx 
\Delta ^{ab}_{\alpha \beta
 }(q,p) \ne 0~~;~~\{\Theta ^a_\alpha (q,p) ,\Psi ^l_\beta (q,p)\}\approx 0
$$
\begin{equation}
 \{\Psi _l(q,p) ,\Psi _n (q,p)\}\approx 0 ~~;~~
\{\Psi _l(q,p) ,H (q,p)\}\approx 0.
\label{1}
\end{equation}
In the above $(q,p)$ are referred to as physical variables and "$\approx $"
means that the equality holds on the constraint surface and
$\{p_a,x_b\}=g_{ab}~~;~~g_{ab}=diag.(1,-1,-1)$. Clearly $\Theta 
_\alpha ^a $ and $\Psi _l $ are SCC and FCC \cite{10} respectively. 
The latter 
are responsible for gauge invariance and the former can be used as
operator identities provided one uses the DBs  \cite{10} as defined below,
\begin{equation}
\{A(q,p),B(q,p)\}_{DB}=\{A,B\}-\{A,\Theta _\alpha ^a\}
\Delta _{ab}^{\alpha \beta }\{\Theta _\beta ^b, B\} ~,~
\Delta ^{ab}_{\alpha \beta }\Delta _{bc}^{\beta \gamma }=\delta ^a_c
\delta ^\alpha _\gamma .
\label{2}
\end{equation}
However, in systems with non-linear SCCs, in general the DBs can become
dynamical variable dependent \cite{5,6,15}
due to the $\{A,\Theta _\alpha ^a\}$ and $\Delta _{ab}^{\alpha \beta }$
 terms, leading to problems for
the quantisation programme. To cure this type of pathology, BT formalism
is a systematic framework where one introduces the BT variables $\phi
 ^\alpha _a $, obeying
\begin{equation}
\{\phi ^\alpha _a,\phi ^\beta _b\}=\omega ^{\alpha \beta}_{ab}=
-\omega ^{\beta \alpha}_{ba},
\label{3}
\end{equation}
where $\omega ^{\alpha \beta}_{ab}$ is a constant (or at most
 a c-number funtion) matrix, with the aim of modifying the SCC 
$\Theta _\alpha ^a(q,p)$ to $\tilde \Theta _\alpha ^a (q,p,\phi ^\alpha _a)$
such that, 
\begin{equation}
\{\tilde\Theta ^a_\alpha (q,p,\phi ) ,\tilde\Theta ^b_\beta (q,p,\phi )\}=0
~~;~~\tilde\Theta ^a_\alpha (q,p,\phi )=\Theta ^a_\alpha (q,p)+
\Sigma _{n=1}^\infty \tilde\Theta ^{a(n)} _\alpha (q,p,\phi )~~;~~
\tilde\Theta ^{a(n)}\approx O(\phi ^n)
\label{4}
\end{equation}
The explicit terms in the above expansion are \cite{12},
\begin{equation}
\tilde\Theta ^{a(1)}_\alpha =X^{ab}_{\alpha \beta }\phi ^\beta _b~~;~~
\Delta ^{ab}_{\alpha \beta }+X^{ac}_{\alpha \gamma }
\omega _{cd}^{\gamma \delta }X^{bd}_{\beta \delta }=0
\label{5}
\end{equation}
 
\begin{equation}
\tilde\Theta ^{a(n+1)}_\alpha =-{1\over{n+2}}
\phi_{d}^{\delta }\omega ^{dc}_{\delta \gamma }X_{cb}^{\gamma \beta }B^{ba(n)}_{\beta \alpha }~~;~~n\ge 1
\label{6}
\end{equation}

\begin{equation}
B^{ba(1)}_{\beta \alpha }=
\{\tilde\theta ^{b(0)} _\beta ,\tilde\theta ^{a(1)} _\alpha \}_{(q,p)}-
\{\tilde\theta ^{a(0)} _\alpha ,\tilde\theta ^{b(1)} _\beta \}_{(q,p)}
\label{7}
\end{equation}

\begin{equation}
B^{ba(n)}_{\beta \alpha }=
\Sigma _{m=0}^n 
\{\tilde\theta ^{b(n-m)} _\beta ,\tilde\theta ^{a(m)} _\alpha \}_{(q,p)}+
\Sigma _{m=0}^n
\{\tilde\theta ^{b(n-m)} _\beta ,\tilde\theta ^{a(m+2)} _\alpha \}_{(\phi )}
~~;~~n\ge 2
\label{8}
\end{equation}
In the above, we have defined,
\begin{equation}
X^{ab}_{\alpha \beta }X_{bc}^{\beta \gamma }=
\omega ^{ab}_{\alpha \beta }\omega _{bc}^{\beta \gamma }
=\delta ^\gamma _\alpha \delta ^a _c .
\label{9}
\end{equation}
A very useful
idea is to introduce the Improved Variable $\tilde f(q,p)$ \cite{12}
 corresponding to each $f(q,p)$,
\begin{equation}
\tilde f(q,p,\phi )\equiv f(\tilde q, \tilde p )
=f(q,p)+\Sigma _{n=1}^\infty \tilde f(q,p,\phi )^{(n)}~~
;~~\tilde f^{(1)}=-
\phi_{c}^{\beta }\omega ^{cb}_{\beta \gamma }X_{bd}^{\gamma \delta }\{
\theta_\delta ^a ,f\}_{(q,p)}
\label{10}
\end{equation}

\begin{equation}
\tilde f^{(n+1)}=-{1\over{n+1}}
\phi_{c}^{\beta }\omega ^{cb}_{\beta \gamma }X_{bd}^{\gamma \delta }
G(f)^{d(n)}_\delta ~~;~~n\ge 1
\label{11}
\end{equation}

\begin{equation}
G(f)^{b(n)}_{\beta }=
\Sigma _{m=0}^n 
\{\tilde\theta ^{b(n-m)} _\beta ,\tilde f^{(m)}\}_{(q,p)}+
\Sigma _{m=0}^{(n-2)}
\{\tilde\theta ^{b(n-m)} _\beta ,\tilde f^{(m+2)}\}_{(\phi )}
+\{\tilde\theta ^{b(n+1)} _\beta ,\tilde f^{(1)}\}_{(\phi )}
\label{12}
\end{equation}
which have the property 
$\{\tilde\Theta ^a_\alpha (q,p,\phi ) ,\tilde f(q,p,\phi )\}=0$. It can be
proved that extensions of the original FCC $\Psi _l $ and Hamiltonian
 $H$ are simply,
\begin{equation}
\tilde \Psi _l=\Psi (\tilde q, \tilde p)~~;~~
\tilde H=H (\tilde q, \tilde p).
\label{13}
\end{equation}
One can also reexpress the converted SCCs as 
$\tilde\Theta ^a_\alpha \equiv \Theta ^a_\alpha (\tilde q,\tilde p)$.
The following identification theorem,
\begin{equation}
\{\tilde A,\tilde B \}=\tilde {\{A,B\}_{DB}}~~;~~
\{\tilde A,\tilde B \}\mid _{\phi =0}=\{A,B \}
 _{DB}~~;~~\tilde 0=0,
\label{14}
\end{equation}
will play a crucial role in our later application. Hence the outcome 
of the BT extension is the closed system of FCCs with the FC
Hamiltonian given below,
\begin{equation}
\{\tilde \Theta _\alpha ^a ,\tilde \Theta _\beta ^b\}= 
\{\tilde \Theta _\alpha ^a ,\tilde \Psi _l\}= 
\{\tilde \Theta _\alpha ^a ,\tilde H\}= 0~~;~~
\{\tilde \Psi _l ,\tilde \Psi _n\}\approx 0 ~;~
\{\tilde \Psi _l ,\tilde H\}\approx 0.
\label{15}
\end{equation}
We will see that due to the non-linearity in the SCCs, the extensions 
 in the improved variables, (and subsequently in the FCCs
 and FC Hamiltonian),
turn out to be infinite series. This type of situation has been encountered
 before \cite{15}.
\vskip .5cm
\begin{center}
{\bf III: Spinning Particle Model revisited }
\end{center}
\vskip .2cm
The SPM proposed by us \cite{6} where an anyon interacts with
a $U(1)$ gauge field
in 2+1-dimensions is given by the Lagrangian,
\begin{equation}
L=(m^2U^a U_a +{{j^2}\over 2}\sigma ^{ab}\sigma _{ab}
+mj\epsilon ^{abc}U_a \sigma _{bc})^{{1\over 2}} + eU_a A^a ,
\label{16}
\end{equation}
where 
$$
U^a ={{dx^a }\over {d\tau }}~~;~~\sigma ^{ab}=
{1\over 2}\epsilon^{abc}\sigma _c=\Lambda _c ^{~a}
{{d\Lambda ^{cb }}\over {d\tau }}~~:~~\Lambda _c^{~a}\Lambda^{cb}
=\Lambda _{~c}^a\Lambda^{bc}=g^{ab}.
$$
Here $(x^a~,~\Lambda ^{ab})$ is a Poincare group element and also 
a set of dynamical
variables of the theory.
The canonical momenta are defined in the following way \cite{5,6},
\begin{equation}
p^a=-{{\partial L}\over {\partial U_a}}\equiv \pi ^a-eA^a ~,~
S^{ab}=-{{\partial L}\over {\partial \sigma_{ab}}}\equiv {1\over 2}
\epsilon ^{abc}S_c.
\label{17}
\end{equation}
The phase space algebra is,
\begin{equation}
\{x_a,x_b\}=0~~;~~\{p_a,x_b\}=g_{ab}~~;~~\{\pi _a,\pi _b\}=eF_{ab}
=e(\partial _aA_b-\partial _bA_a),
\label{18}
\end{equation}
\begin{equation}
\{S^a,S^b\}=\epsilon^{abc}S_c~~;~~\{S^a,\Lambda ^{0b}\}=\epsilon^{abc}
\Lambda ^0_{~c}~~;~~\{\Lambda ^{0a},\Lambda ^{0b}\}=0.
\label{19}
\end{equation}
In the free theory, one encounters the following set of FCC and SCC
respectively,
$$
\Psi _1\equiv \pi _a\pi ^a-m^2 \approx 0~~;~~
\Psi _2 \equiv \pi .S-{{mj}\over 2} \approx 0,$$
$$
\Theta ^a_1\equiv \Lambda ^{0a}-{{p ^a}
\over m} \approx 0~~;~~ 
\theta ^a_2\equiv S^{ab}p_b = \epsilon ^{abc}p _b S_c \approx 0 .
$$
In the free theory, $\Psi _1$ and $\Psi _2$ are respectivly the mass-shell
 condition and the Pauli-Lubanski relation.

In the interacting theory,
 One obtains to $O(e)$ the following set of FCCs and SCCs ,
\begin{equation}
\Psi _1\equiv \pi _a\pi ^a-m^2 +{{2eF_{ab}\pi ^a}\over
{\pi ^2}}\Theta _2^b
\approx 0~~;~~\Psi _2 \equiv \pi .S-{{mj}\over 2}
+{{ejF_{ab}\pi ^a}\over
{2m\pi ^2}}\Theta _2^b
\approx 0 ,
\label{20}
\end{equation}
\begin{equation}
\Theta ^a_1\equiv \Lambda ^{0a}-{{\pi ^a}
\over m} \approx 0~~;~~ 
\theta ^a_2\equiv S^{ab}\pi _b = \epsilon ^{abc}\pi _b S_c \approx 0 .
\label{21}
\end{equation}
Actually one can check that $S ^aS _a={{j^2}\over 4} $
but this is not independent of $\Psi _l $ given above (\ref{20}).
To verify the constraint algebra after computing the PBs, one has
to invoke the relation $S^a={j\over {2m}}\pi ^a$, which is consistent
 with the SCC $\Theta ^a _2$ and the normalisation agrees with the free
  theory. The above relation is valid also at the level of DBs
   \cite{6}.

Note that in the above set of SCC, $\Theta ^a_1 $ has been imposed from  
outside such that the angular coordinates are properly restricted. 
It is easy to ascertain that the SCCs $\Theta _\alpha ^a $ form a 
reducible set, since
$$\pi ^a \Theta _{1a}=-{1\over 2}m\Theta ^{1a}\Theta _{1a}~~;~~
\pi ^a\Theta _{2a}=0.
$$
(With $e=0$, these features are true for the free theory stated above.)
However, one works with this reducible set because the manifestly 
covariant structure simplifies calculations and one can invert 
\cite {5,6} the SCC
algebra matrix perturbativly to get the DBs. To $O(e)$ we obtain the DB between two
generic field as \cite{6},
$$
\{A,B\}_{DB}=\{A,B\}+{{eF_{ab}}\over 
{m^2}}\{A,\Theta _2^a\}\{\Theta _2^b,B\}
$$
$$
-{1\over 2}\epsilon ^{abc}S_c\{A,\Theta _{1a}\}\{\Theta _{1b},B\}
+{1\over {2m}}\{A,\Theta _2^a\}\{\Theta _{1a},B\}
$$
\begin{equation}
-{1\over {2m}}\{A,\Theta _1^{a}\}\{\Theta _{2a},B\}.
\label{22}
\end{equation}
In particular the following DB
\begin{equation}
\{x^a ,x^b\}_{DB}= -{1\over {2m^2}}\epsilon ^{abc}S_c +O(e),
\label{23}
\end{equation}
gives rise to the non-commutative space-time,
which in turn generates the arbitrary spin contribution in the angular
momentum. 
Note that the above algebra (\ref{23}) is non-trivial in the free theory as well.
Fixing the gauge condition $x_0=\tau $ as the proper time and
using $\Psi _1 $ one ends up with the Hamiltonian,
\begin{equation}
H=(m^2-\pi ^i \pi _i )^{{1\over 2}} -eA_0,
\label{24}
\end{equation}
where the $\Theta _2 ^a$ dependent term has been dropped since we
 used DBs \cite{6,5}. To $O(e)$ 
the 2+1-dimensional analogue of the 3+1-dimensional Bargmann-Michel-
Telegdi equation \cite{16} is,
\begin{equation}
\dot S^a=\{H,S^a\}_{DB}={e\over m}F^{ab}S_b .
\label{25}
\end{equation}
Comparison with the original equation \cite{16}
 reveals that $g=2$ for the particle.

As we have mentioned before, the comparison has been carried out between
the present system in non-commutative  configuration space and the original
 Bargmann-Michel-Telegdi 
equation in normal coordinate system.

The connection between our result and that of SM \cite{4} 
is very direct but subtle. Remember that in SM \cite{4}, a modified
expression, (with an $O(eF)$ term), for the "mass-shell" condition is 
used. (In our analysis, an analogous
 $\Theta ^a_2$-dependent term  in (\ref{24}) was
"strongly" put to zero since it
was proportional to the SCCs and we use DBs.) The analysis in SM is for
 FCC system, according to Dirac, where one demands that the FCCs 
annihilate the physical states. (In our BT extended theory, we will also
 follow this route.) Now in \cite{4}, one solves perturbativly,
 for small $e$ the non-canonical symplectic algebra, (which is equivalent
 to
our DBs), in terms of a set of canonical phase space variables and
rewrites the Hamiltonian in terms of these {\it new} 
variables and finally
compares this expression with the Hamiltonian of a charged particle in 
ordinary space time. Again the $g=2$ result \cite{4} is reproduced. 
Obviously this derivation \cite{4} also suffers from the 
same  conceptual drawback as the previous 
one \cite{6}.
 With this
background, we now move on to the BT extension of SPM.

\vskip .5cm
\begin{center}
{\bf IV:Batalin-Tyutin extension of Spinning Particle Model}
\end{center}
\vskip .2cm
This section comprises of the main body of our work where 
we introduce the BT machinary \cite{12}
in order to take into account the SCCs of the theory but at the same 
time avoid using
a non-canonical coordinates. As mentioned before, we now 
use a smaller set of SCCs which are
irreducible. Also note that in the extended space the 
constraints will be
modified and we can no longer use the original constraints 
to simplfy the SCC
algebra. Choosing $\Theta _\alpha ^i$ as the irreducible 
set of SCCs, the PB algebra is,
\begin{equation}
\{\Theta ^i_\alpha ,\Theta ^j_\beta \}=\Delta ^{ij}_{\alpha \beta }
\label{26}
\end{equation}
\begin{equation}
\Delta ^{ij}_{11}={{eF^{ij}}\over{m^2}}
\label{27}
\end{equation}

\begin{equation}
\Delta ^{ij}_{12}=-[(\pi .\Lambda )g^{ij}-\pi ^i\Lambda ^{0j}+{e\over m}
\epsilon ^{jbc}F^i_{~b}S_c ]=-\Delta ^{ji}_{21}~~;~~\pi .\Lambda =\pi _a
\Lambda ^{0a}
\label{28}
\end{equation}

\begin{equation}
\Delta ^{ij}_{22}=[\pi _0(\pi .S)\epsilon ^{ij}+e(F^{ij}S^2+S_bF^{bi}S^j
+S^iF^{jb}S_b]~~;~~S^2=S_aS^a .
\label{29}
\end{equation}
Next, following (\ref {5}),  we need to compute $X_{\alpha\beta }^{ij}$
 as the whole calculational scheme rests on this quantity and its
  inverse. This
we do perturbativly by first considering the free theory, (i.e. $e=0$),
 where,

\begin{equation}
\Theta ^i_1\equiv \Lambda ^{0i}-{{p^i}
\over m}~~;~~\Theta ^i_2\equiv \epsilon ^{ibc}p_b S_c
\label{30}
\end{equation}

\begin{equation}
\{\Theta ^i_\alpha ,\Theta ^j_\beta \}=\Delta ^{ij}_{\alpha \beta }
=
\left (
\begin{array}{cc}
0 & -((p.\Lambda )g^{ij}-p^i\Lambda ^{0j})\\
((p.\Lambda )g^{ij}-p^j\Lambda ^{0i}) & p_0(p.S)\epsilon ^{ij}
\end{array} \right ).
\label{m1}
\end{equation}
For the free theory we propose,
\begin{equation}
X_{\alpha \beta }^{ij}\mid _{e=0}\equiv x_{\alpha \beta }^{ij}=
\left (
\begin{array}{cc}
0 & -((p.\Lambda )g^{ij}-p^i\Lambda ^{0j})\\
g^{ij} & {1\over 2}p_0(p.S)\epsilon ^{ij}
\end{array} \right ). 
\label{m2}
\end{equation}
One can check that $x_{\alpha \beta }^{ij}$ satisfies (\ref{5}) for the free theory,
 provided we choose 
$\omega ^{\alpha \beta }_{ab}=\epsilon ^{\alpha \beta }g_{ab}~,~
\epsilon ^{12}=1$. Since 
there are some artitrariness
 involved in $x_{\alpha \beta }^{ij}$ and
 $\omega ^{\alpha \beta }_{ab}$, their choices are dictated by
 convenience. The inverse is defined below,
$$
x^{\alpha \beta }_{ij}x_{\beta \gamma }^{jk}=\delta ^\alpha _\gamma g^i_k,
$$
 \begin{equation}
x^{\gamma \delta }_{jk}=
 \left (
\begin{array}{cc}
{{p_0(p.S)}\over {2(p.\Lambda )}}(\epsilon _{jk}-{{\epsilon _
{jl}p_l\Lambda _{0k}}
\over {p_0\Lambda _{00}}}) & g_{jk}\\
-{1\over{(p.\Lambda )}}(g_{jk}+{{p_j\Lambda _{0k}}\over{p_0\Lambda _{00}}})
& 0
\end{array}  \right ).
\label{m3}
\end{equation}
This is an exact result.
Now, for the interacting theory, we find to $ O(e)$,
\begin{equation}
X_{\alpha \beta }^{ij}=x_{\alpha \beta }^{ij}(p_a\rightarrow \pi _a)+
ey_{\alpha \beta }^{ij}
\label{31}
\end{equation}

\begin{equation}
y_{11}^{ij}={{F^{il}\Lambda _{0l}p^j}\over{m^2p_0
\Lambda _{00}(p^k\Lambda _{0k})}}
~~;~~y_{22}^{ij}={1\over 2}(S_0)^2F^{ij}
\label{IV}
\end{equation}

\begin{equation}
y_{21}^{ij}={{S_0\epsilon ^{rl}F^{0r}S^l}\over{p_0(p.S)}}g^{ij}
\label{32}
\end{equation}

\begin{equation}
y_{12}^{ij}=-{1\over m}F^i_{~b}\epsilon ^{jbc}S_c+
{{F^{il}\Lambda _{0l}p_k\epsilon^{jk}p_0(p.S)}
\over{2m^2p_0\Lambda _{00}(p^k\Lambda _{0k}}}+
{{S_0\epsilon ^{rl}F^{0r}S^l}\over{p_0(p.S)}}((p.\Lambda )g^{ij}
-p^i\Lambda ^{0j})
\label{33}
\end{equation}
The inverse, to $O(e)$, is  of the form, 
\begin{equation}
X^{\alpha \beta }_{ij}=x^{\alpha \beta }_{ij}(p_a\rightarrow \pi _a)
-ex^{\alpha \mu }_{ik}(p)y_{\mu \nu }^{kl}(p)x^{\nu \beta }_{lj}(p).
\label{34}
\end{equation}
In the present work the explicit form of $X^{\alpha \beta }_{ij}$ will 
not be utilised.
From the relation,
\begin{equation}
\tilde\Theta ^{i(1)}_\alpha =X_{\alpha \beta }^{ij}\phi _j^\beta ,
\label{35}
\end{equation}
the explicit expressions for $\tilde\Theta ^{i(1)}_\alpha $ (i.e. one BT variable
extension term) are,
\begin{equation}
\tilde\Theta ^{i(1)}_1=-(g^{ik}(\pi . \Lambda )-\pi ^i\Lambda ^{0k})\phi ^2_k
+e(y^{ik}_{12}\phi^2_k+y^{ik}_{11}\phi^1_k),
\label{36}
\end{equation}
\begin{equation}
\tilde\Theta ^{i(1)}_2={1\over 2}\pi _0(\pi .S)\epsilon ^{ik}\phi ^2_k +\phi ^{1i}
+e(y^{ik}_{21}\phi^1_k+y^{ik}_{22}\phi^2_k),
\label{37}
\end{equation}

We emphasise that the series for $\tilde\Theta ^{i}_1$ and $\tilde\theta ^{i}_2$
do not terminate and the higher order terms in $\phi ^\alpha _i$ can be derived
 by  a straightforward but extremely tedious calculation. 
However, to check whether our target of converting SCCs to FCCs has really been
 achieved (to $O(e)$),
 one can convince oneself that,
$$
\{(\Theta _\alpha ^i+\tilde \Theta _\alpha ^{i(1)})~,~
(\Theta _\beta ^j+\tilde \Theta _\beta ^{j(1)})\}=0+O(\phi ).
$$
To check the cancellation of $O(\phi )$ terms, $\Theta _\alpha ^{i(2)}$
 terms are required.

Now we move on to the $one~ BT ~extension$ of the 
physical degrees of freedom,  i.e the Improved 
Variables. The BT extensions of the BT variables themselves will vanish.
Let us start by constructing the extensions for $\pi ^a$. From the
generic expression given in (\ref{10}), we have,
$$\tilde \pi ^{a(1)}=-
\phi_{c}^{\beta }\omega ^{cb}_{\beta \gamma }X_{bd}^{\gamma \delta }\{
\Theta_\delta ^d ,\pi ^a\}_{(q,p)}
$$
$$
={e\over m}(\phi ^{2j}x^{11}_{ji}-\phi ^{1j}x^{21}_{ji})F^{ia}-
e\phi ^2_i\epsilon ^{ide}S_eF_d^{~a}
$$
$$
={e\over m}F^{ia}[(\phi ^{2j}{{p_0(p.S)}\over{2(p.\Lambda)}}
(\epsilon _{ji}-{{\epsilon _{jl}p_l\Lambda ^0_{~i}}
\over{p_0\Lambda _{00}}})
$$
\begin{equation}
+\phi ^{ij}{1\over {(p.\Lambda )}}
(g_{ij}+{{p_j\Lambda _{0i}}\over {p_0\Lambda _{00})}})]-
e\phi ^2_i\epsilon ^{ide}S_eF_d^{~a}.
\label{37a}
\end{equation}
In a straightforward manner, one can compute the rest of the Improved
 Variables, from the structures given below,
$$
\tilde x^{a(1)}=
-\phi_{c}^{\beta }\omega ^{cb}_{\beta \gamma }X_{bd}^{\gamma \delta }\{
\Theta_\delta ^d ,x^a\}_{(q,p)}
$$
\begin{equation}
={{g^{ai}}\over m}
(\phi ^{2j}x^{11}_{ji}-\phi ^{1j}x^{21}_{ji})+\epsilon ^{aic}S_c
(\phi ^{2j}X^{12}_{ji}-\phi ^{1j}X^{22}_{ji}),
\label{38}
\end{equation}

$$\tilde S^{a(1)}=
-\phi_{c}^{\beta }\omega ^{cb}_{\beta \gamma }X_{bd}^{\gamma \delta }\{
\Theta_\delta ^d ,S^a\}_{(q,p)}
$$
\begin{equation}
=\epsilon ^{aic}\Lambda _{0c}
(-\phi ^{1j}X^{21}_{ji}+\phi ^{2j}X^{11}_{ji})-(g^{ai}(\pi .S)-S^i\pi ^a)
(\phi ^{2j}X^{12}_{ji}-\phi ^{1j}X^{22}_{ji})
\label{39}
\end{equation}

$$
\tilde \Lambda ^{0a(1)}=
-\phi_{c}^{\beta }\omega ^{cb}_{\beta \gamma }X_{bd}^{\gamma \delta }\{
\Theta_\delta ^d ,\Lambda ^{0a}\}_{(q,p)}
$$
\begin{equation}
=(g^{ai}(\pi .\Lambda )-\Lambda ^{0i}\pi ^a)
(\phi ^{1j}X^{22}_{ji}-\phi ^{2j}X^{12}_{ji}).
\label{40}
\end{equation}
These Improved Variables also comprise of 
infinite sequences of higher order $\phi ^\alpha _i$ 
terms. As a non-trivial consistency check, we have tested the validity
of the assertion that, to $O(\phi )$, 
$\tilde\Theta ^a_\alpha \equiv \Theta ^a_\alpha (\tilde q,\tilde p)$ 
holds. To examine  the $two-\phi $-term, $O(\phi ^2)$-terms in $\tilde\Theta 
^i_\alpha $ and $\tilde f(q,p)$ are required.  
In the next section we will make use of these results to 
redetermine the $g$-value
for the anyon in the extended phase space. However, as we have
 stressed before, the main achievement is that a consistent
framework has been provided wherein quantisation of the SPM
is unambiguous and derivation of the previous results are more
 transparant.

\vskip .5cm
\begin{center}
{\bf V: Application - $g$ in extended space}
\end{center}
\vskip .2cm

Let us start by recovering the "mass shell" condition $\tilde\Psi _1 $ in
BT extended space, which is simply given by
\begin{equation}
\tilde\Psi _1\equiv \tilde\pi _a\tilde\pi ^a-m^2 +{{2e\tilde F_{ab}
\tilde\pi ^a}\over
{\tilde\pi ^2}}
\tilde\Theta _2^b
\approx 0.
\label{41}
\end{equation}
Remembering that {\it all} the Improved Variables are of the form 
$$\tilde{\cal A}(q,p,\phi )={\cal A}(q,p) +O(\phi )
~~+..~~;~~ \tilde F=F(\tilde x)
=F(x)+(\partial F)\phi +~~..~~, $$
and neglecting $\partial F$ terms, 
it is clear that $\phi $-independent {\it polynomial} expressions of 
 relevant operators will remain intact and so up to non-$\phi $
terms, the previous relations survive. Hence proceeding in the 
same way as before, in the extended space, one can compare the
Schrodinger equation to derive $g=2+O(\phi )$ for anyons. However,
more work is to be done to ascertain whether the $\phi $-terms can
contribute to $g$ even in the non-relativistic limit $p_0
\approx m>>\mid {\bf p}\mid $
used in \cite{4}. Notice that the Pauli-Lubanski relation
is modified to, 
$$\tilde\Psi _2 \equiv 
\tilde\pi .\tilde S-{{mj}\over 2} 
+{{ej\tilde F_{ab}
\tilde\pi ^a}\over
{2m\tilde\pi ^2}}
\tilde\Theta _2^b
\approx 0.$$ 
This discussion corresponds to the SM model \cite{4}. 

Returning to our SPM model \cite{6}, we follow
 the discussion in Section III and introduce the gauge fixing condition for 
$\tilde\Psi _1 $ 
to be $\tilde x_0-\tau = 0$ and obtain the Hamiltonian as,
\begin{equation}
\tilde H=(m^2-\tilde\pi ^i\tilde \pi _i+\tilde \Theta _2^a-term )
^{{1\over 2}} -e\tilde A_0.
\label{42}
\end{equation}
We immedietly notice that
the Hamiltonian in extended space
is not a rational function of the phase space variables and also that 
the $\Theta _2^a $ term can not be dropped. But the theorem in
 (\ref{14}) comes to the rescue. The $\Theta _2^a$ can contribute
to the equations of motion of the generic form $\dot {\tilde{\cal A}}=
\{\tilde H,\tilde{\cal A}\}$ by terms proportional 
to $\tilde\Theta _2^a$ only since since PBs in extended space with
$\tilde\Theta _\alpha ^i$ vanish. So long as we are concerned with
computing PBs, the theorem (\ref{14}) can be applied without going to the 
details of the explicit structure of the respective operators in the PB.
 Hence utilising (\ref{14}),
 we can simply transform the BMT
equation of motion for the spin variable \cite{6}, in terms of the 
Improved Variables and get,
\begin{equation}
\dot {\tilde S}^a={e\over m}\tilde F^{ab}\tilde S_b \approx
{e\over m}F^{ab}\tilde S_b .
\label{43}
\end{equation}
This is our cherished Bargmann-Michel-Telegdi equation for anyon in BT-
extended space. 
Obviously in  the the BT extended space, the
expression for $g$ is modified to 
$g=2+O(\phi )$ for anyons, with the additional terms coming from the BT-
variable dependent terms.

To facilitate a comparison with the results obtained in the physical
 variable sector, (without introducing BT variables), we can rewrite
 the Improved Variables by their physical counterpart and the BT extension
  terms. Since the phase space PB algebra (\ref{18},\ref{19},
  \ref{3}) is trivially
   known, we can compute $\dot S^a$ from (\ref{45}) in a
    straightforward way. Obviously it will be of the form,
    \begin{equation}
\dot {\tilde S}^a \approx
{e\over m}F^{ab}\tilde S_b  +O(\phi ).
\label{bmt}
\end{equation}

In a restricted way, we can eliminate the BT variables in terms of the physical
variables if we adopt the Dirac quantisation prescription for systems with
FCCs only, where the physical sector of Hilbert space
is required to satisfy $FCC\mid Physical~ State
\rangle =0$ \cite{17}. In the present case,
this reduces to
\begin{equation}
\tilde \Theta _\alpha ^i \mid Ph. ~St.\rangle =\tilde \Psi _l\mid Ph.~St.
\rangle =0.
\label{44}
\end{equation}
Since we are interested in substituting the BT variables in the expression of the
Hamiltonian in (\ref{42}), we can solve the constraints for $e=0$ as the terms involving BT
variables are already of $O(e)$. This leads to the set of equations,
$$
\tilde \Theta _1 ^i\mid _{e=0}
=\Theta _1^i\mid _{e=0}-(g^{ij}(p.\Lambda )-p^i\Lambda ^{oj})
\phi ^2_j=0, 
$$
\begin{equation}
\tilde \Theta _2 ^i\mid _{e=0}
=\Theta _2^i\mid _{e=0}+\phi ^{1i}+{1\over 2}m^2S_0\epsilon
^{ik}\phi ^2_k=0.
\label{45}
\end{equation}
 
From the solutions of the above equations, we find,
$$
\phi ^{1i}=-\Theta _2^i-{1\over 2}m^2S_0\epsilon^{ij}\phi ^2_j,
$$
\begin{equation}
\phi ^{2i}={{\Theta _{1j}}\over {(p.\Lambda )}}(g^{ij}+{{p^i\Lambda ^{0j}}\over
{p_0\Lambda _{00}}})
\label{46}
\end{equation}
Notice that 
the BT-variables in this restricted derivation, being proportional to
$\Theta _\alpha ^i$, are non-vanishing even in the free theory since in 
the extended sector $\Theta _\alpha ^i$ are no longer the constraints.
It should be emphasised that this is not the whole story since $\tilde
\Theta_\alpha ^i $ consists of an infinite number of terms. Also there are 
indications that higher order terms will survive even if, on top of our small
$e$ restriction, we incorporate the non-relativistic limit, (that is 
$p_0\approx m>>\mid {\bf p }\mid $, as has been done in \cite{4}). On the
other hand, putting these expressions in the BMT equation shows the leading
term in $g=2$ survives but there can be non-trivial velocity 
dependent corrections. Putting the BT variables back in (\ref{bmt}) we
get the BMT equation completely in the physical sector.

\vskip .5cm
\begin{center}
{\bf VI: Conclusion}
\end{center}
\vskip .2cm

To conclude, in the present work we have formulated an extension of
the Spinning Particle Model for anyon \cite{6} along the lines of
Batalin-Tyutin quantisation scheme \cite{12}. The reason behind working
 in the extended phase space lies in the presence of non-linear Second
 Class Constraints in the model, which induce a non-canonical structure
 in the particle coordinate. Specifically, the position coordinates become
 non-commuting which creates problem for the quantisation programme. Also this 
makes the comparison between results obtained here and in conventional 
space-time models, difficult. The gyromagnetic ratio of anyon was obtained
 to be 2 \cite{4,6} by invoking precisly this type of matching. 

To avoid this type
of non-trivial structure in the space-time, the Batalin-Tyutin formalism
 \cite{12} is adopted where extra BT-variables are introduced in the
phase space in such a way that all the constraints become First Class in the
extended space and the problematic Dirac Brackets can be avoided. Hence the
commuting space-time structure is kept intact. However, the price to pay is
that the extensions of the constraints and relevant variables turn out to
be infinite serieses, (even to lowest order in $e$, the
electromagnetic coupling), with higher powers of BT variables. One has to be
very careful in taking the non-relativistic limit \cite{4} and {\it a priory}
it is difficult to determine whether the serises will terminate or not. 

In
the present work we have computed explicitly the one-$\phi $ extensions of
{\it all} the constraints and degrees of freedom. From the nature of the
 extensions, it is clear the BT-variable independent relations remain the same
as the original ones. Hence one might say that the BT terms are effects of the
non-commuting space-time algebra.  
 However, 
we note that these results are  partial in the sense that higher BT-variable terms
may also contribute to this order of accuracy. One of our immediate goals in
this area is to compute explicitly effect of the BT variable terms in
 observable quantities, {\it e.g.} $g$.

From another point of view, this work is significant since it may provide
 a mapping between theories in {\it non-commuting} space-time on one hand, and
theories in conventional configuration space with extra {\it spin} 
degrees of freedom. Subsequently, Batalin-Tyutin extension can be introduced
and one can check if results of simple non-commutative models \cite{9} are
 reproduced.

Another interesting problem of the formal kind is to develop the 
Batalin-Tyutin scheme for reducible Second Class Constraint systems, such
as the Spinning Particle Model in manifestly covariant form. One has to
be careful in introducing the BT variables since 
reducibility in the system will be reflected in the number of these
degrees of freedom.

\end{document}